\documentclass[10pt,twocolumn,letterpaper]{article}

\usepackage{cvpr}              

\usepackage[dvipsnames]{xcolor}

\definecolor{cvprblue}{rgb}{0.21,0.49,0.74}
\usepackage[pagebackref,breaklinks,colorlinks,citecolor=cvprblue]{hyperref}

\def\paperID{6} 
\def\confName{FedVision}
\def\confYear{2024}

\usepackage{soul}
\usepackage{url}
\usepackage{graphicx}
\usepackage{amsmath}
\usepackage{amsthm}
\usepackage{booktabs}
\usepackage{algorithm}
\usepackage{algorithmic}
\usepackage[switch]{lineno}
\usepackage{physics}
\usepackage{multicol}
\usepackage{array}
\usepackage{multirow}
\usepackage{subcaption}

\title{On the Efficiency of Privacy Attacks in Federated Learning}

\author{
Nawrin Tabassum$^{1}$, Ka-Ho Chow$^2$, Xuyu Wang$^3$, Wenbin Zhang$^3$, Yanzhao Wu$^3$\thanks{Corresponding author.}\\
$^1$ Ahsanullah University of Science and Technology\\
$^2$ University of Hong Kong\\
$^3$ Florida International University\\
{\tt\small nawrin.cse@aust.edu, kachow@cs.hku.hk, \{xuywang, wenbin.zhang, yawu\}@fiu.edu}
}

\begin{document}
\maketitle

\begin{abstract}

Recent studies have revealed severe privacy risks in federated learning, represented by Gradient Leakage Attacks. However, existing studies mainly aim at increasing the privacy attack success rate and overlook the high computation costs for recovering private data, making the privacy attack impractical in real applications. In this study, we examine privacy attacks from the perspective of efficiency and propose a framework for improving the Efficiency of Privacy Attacks in Federated Learning (EPAFL). We make three novel contributions. First, we systematically evaluate the computational costs for representative privacy attacks in federated learning, which exhibits a high potential to optimize efficiency. Second, we propose three early-stopping techniques to effectively reduce the computational costs of these privacy attacks. Third, we perform experiments on benchmark datasets and show that our proposed method can significantly reduce computational costs and maintain comparable attack success rates for state-of-the-art privacy attacks in federated learning. We provide the codes on GitHub at \url{https://github.com/mlsysx/EPAFL}.

\end{abstract}

\section{Introduction}

Federated Learning (FL) follows a decentralized approach to train Machine Learning (ML) models by transmitting gradients and avoiding sharing private user data. 
These methods have gained huge popularity in recent years in the computer vision domain due to their privacy-preserving capabilities for training vision models. They are especially useful in cases where training data is highly sensitive and should not be shared over the internet, e.g., healthcare \cite{jochems2017developing,sheller2020federated} and finance \cite{wen2023survey,mammen2021federated} data.
In practice, FL models~\cite{mcmahan2017communication,konevcny2016federated} are trained on edge devices (e.g., smartphones, IoT devices), where only the gradients are uploaded to the central server and aggregated to update the global model. The updated global model will be sent back to the client devices for the next round of training. 
However, several recent studies have revealed that FL is vulnerable to various security risks and privacy attacks, such as data poisoning attack~\cite{tolpegin2020data,wei2024demystifying}, model poisoning attack \cite{fang2020local}, model hijacking~\cite{salem2022get,chow2023STDLens}, GAN-based attack \cite{zhang2020poisongan}, backdoor attack \cite{bagdasaryan2020backdoor}, and gradient leakage attack~\cite{zhu2019deep}.
In particular, the gradient leakage attack has received growing attention~\cite{zhu2019deep,zhao2020idlg,wei2020framework}.
Though training data are not shared directly in FL, the input data and labels can still be extracted from the shared gradients. This phenomenon is known as the gradient leakage attack, which puts the privacy guarantee of FL into question. 
The gradient leakage attacks initialize dummy data and calculate the dummy gradients from these data. Then the distance between actual and dummy gradients is minimized through an iterative process to update the dummy data so that the dummy data will gradually become similar to the private data.
Following this approach, the gradient leakage attack was first proposed in~\cite{zhu2019deep} to reconstruct training data from shared gradients. Subsequently, an improved gradient attack was shown in \cite{zhao2020idlg} where both data and labels can be reconstructed with very high accuracy. 

\begin{figure}
  \includegraphics[width=\linewidth]{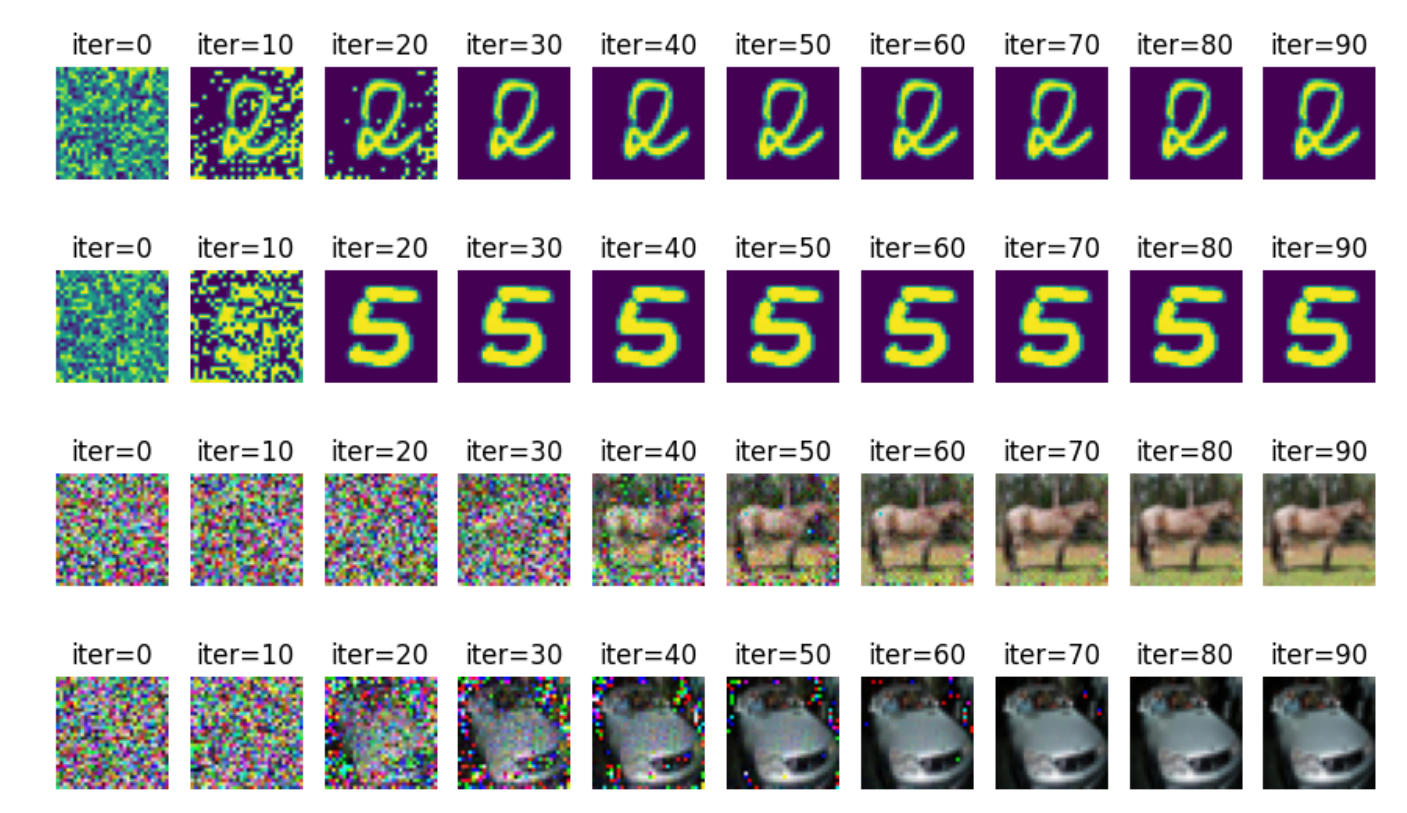}
  \caption{Visualization of Privacy Attack}
  \label{fig:attack-visualize}
  \vspace{-5mm}
\end{figure}

Most existing studies \cite{wei2023securing,wei2023model,wei2020framework,zhu2019deep,zhao2020idlg} on gradient leakage attacks mainly focus on achieving a higher Attack Success Rate (ASR) and/or better reconstruction quality of private training data. Gradient leakage attacks primarily leverage a pre-defined fixed number of iterations to recover private training data, lacking the adaptability to improve the privacy attack efficiency. For example, Figure~\ref{fig:attack-visualize} visualizes the privacy attack process on two image examples from MNIST~\cite{mnistlenet} (first two rows) and two image examples from CIFAR-10~\cite{cifar10-100} (bottom two rows) datasets for 100 iterations. We observe that different numbers of iterations are required to successfully recover different images. It incurs unnecessary computational costs to perform privacy attacks for the entire pre-defined number of iterations, potentially rendering these attacks ineffective under time constraints. On the other hand, efficient privacy attacks can also facilitate the quick design and evaluation of defense strategies. However, very few studies have studied the efficiency challenges of privacy attacks. Several open challenges remain to be addressed: (1) how to optimize the privacy attack efficiency without compromising the attack effectiveness and (2) how to determine the adequate number of iterations required to successfully recover private training data.

In this paper, we present a holistic framework to improve the Efficiency of Privacy Attacks in Federated Learning (EPAFL).
{\it First}, we examine the computational costs of various privacy attacks to identify prospects for improving the efficiency of these attacks.
{\it Second}, we provide different early stopping techniques in EPAFL to dynamically adjust the number of iterations required for reconstructing private data, which significantly reduces the data reconstruction costs.
{\it Third}, we perform comprehensive experiments on two benchmark datasets (MNIST~\cite{mnistlenet} and CIFAR-10~\cite{cifar10-100}), which demonstrates that EPAFL can effectively optimize the efficiency of privacy attacks.

\section{Related Work}

\subsection{Federated Learning}
Traditional centralized machine learning typically trains models on a single server using private data. This practice not only entails high computation requirements of the server but also poses high risks of leaking private data. With the increasing size and cost of deep learning models and the introduction of stringent privacy regulations, distributed machine learning~\cite{dean2012large,kim2016deepspark,sergeev2018horovod,li2020federated,mcmahan2017communication,konevcny2016federated,wu2021parallel,wu2022comparative}, represented by Federated Learning (FL)~\cite{li2020federated,mcmahan2017communication,konevcny2016federated}, emerged to meet these demands, where the computational power of multiple computing nodes (e.g., edge devices) are utilized for training a global ML model. In FL, the local models are trained on client devices using private data and only the gradients are uploaded to the server. Upon receiving these gradients, the server aggregates them and updates the global model following an aggregation algorithm \cite{li2019convergence,pillutla2022robust,reddi2020adaptive,liagg2020federated,wang2020federated}. 
The updated global model is then distributed to the client devices to continue the next round of local training. As training data is not directly shared with the server, FL preserves the confidentiality of sensitive private data. This approach has been adopted in several real-world cases to protect data privacy, such as building predictive models for medical diagnosis~\cite{jochems2016distributed,das2023privacy}, analyzing customer data for credit card fraud detection~\cite{yang2019ffd}, and training personalized models for next-word prediction in keyboards \cite{hard2018federated}.

\subsection{Privacy Attacks in Federated Learning}
Recent studies have investigated privacy risks in FL and showed that the privacy of training data can be easily compromised through shared gradients~\cite{zhu2019deep,zhao2020idlg,wei2020framework}. Gradient leakage attack is first proposed in \cite{zhu2019deep} to recover private training data by minimizing the $L_{2}$ distance between actual and dummy gradients. An improved gradient leakage attack is introduced in~\cite{zhao2020idlg} where labels are also reconstructed with very high accuracy. 
Another privacy attack is proposed in \cite{zhang2020poisongan} where the input images are initialized using GAN models to keep dummy inputs closer to the actual inputs. 
The generalizability of these privacy attacks can be improved by utilizing different distance loss functions~\cite{geiping2020inverting,wang2020sapag}. For example, \cite{geiping2020inverting} leverages the cosine similarity to measure the gradient difference between actual and dummy data, and \cite{wang2020sapag} introduces a weighted Gaussian kernel-based distance function. 
A few studies have made efforts to design new loss functions, such as adding regularization terms~\cite{wei2020framework, yin2021see}. 
In addition, a recursive gradient attack is proposed in \cite{zhu2020rgap}, which performs better than the optimization-based attacks.
Though most of the aforementioned studies use a single input to demonstrate privacy attacks, a few studies also suggest methods for the reconstruction of batch data~\cite{yin2021see, jin2021cafe}, which enhance the single-input attacks and show that private training data can be recovered even after utilizing a very large batch size.

These existing studies have been focused on increasing the attack success rate for improving the gradient leakage attacks.
None of these studies have evaluated the attack performance in terms of the reconstruction costs. Although these privacy attack methods can maintain high reconstruction quality, we observe that they can be further optimized by reducing the computational costs. 

\section{Problem Statement}

\subsection{Federated Learning}

The objective of Federated Learning (FL) is to train a model in a distributed manner without sharing the private training data, where only the gradients are transmitted to the central server \cite{mcmahan2017communication}. 
A global model is built by the server from the gradients shared by the clients. At round $r$, the server first sends the global model $F_{W}^r$ to a randomly selected subset of clients $K$. Each of these clients has its own local dataset $\{(X_{k}, Y_{k})\}_{k=1}^K$. The clients calculate their local gradients $\nabla_{W_k} L$ of the loss function $L$ on their local dataset $(X_{k}, Y_{k})$, update their local models $F_{W_k}$, and send the gradients to the server. Finally, the server aggregates the gradients and updates the global model for the next round of FL training as Formula~\ref{formula:fedsgd} shows.
\begin{equation}
F_{W}^{r+1} = F_{W}^r - \alpha\sum_{k=1}^K\frac{n_{k}}{n} \nabla_{W_k} L
\label{formula:fedsgd}
\end{equation}
where $\alpha$ is the global learning rate, $n_{k}$ represents the number of samples in $(X_{k}, Y_{k})$ from client $k$, and $n$ is the total number of samples from all the selected clients.

\subsection{Gradient Leakage Attack}
\label{GLA}
Federated Learning may not guarantee data privacy even without sharing private training data. The gradient leakage attack is an optimization-based privacy attack that can fully reconstruct private training data from shared gradients~\cite{zhu2019deep,zhao2020idlg,wei2020framework}. After receiving the gradients, the adversary (e.g., an honest-but-curious server) can initialize dummy data, and match the dummy data to the real private training data by minimizing the distance between real and dummy gradients. The adversary minimizes the following objective (Formula~\ref{formula:L2-distance-privacy-attack}) to closely approach the actual private data $(x,y)$ with the dummy data $(x',y')$.
\begin{equation}
\underset{x',y'}{min \:} ||\nabla_{W} L(F_{W}(x'),y') - \nabla_{W} L(F_{W}(x),y)||^2
\label{formula:L2-distance-privacy-attack}
\end{equation}

During the attack, the dummy gradients are optimized through an iterative process to minimize the loss, which measures the gradient difference between the dummy and private data. Our EPAFL framework can effectively monitor the loss and adapt the actual number of iterations $I$ for performing privacy attacks, which can effectively reduce the total execution time cost.

\subsection{Threat Model}

The goal of the gradient leakage attack is to steal the private training data of the clients from the gradients. We assume that the server is honest in following the FL protocols but curious to access clients' private training data.
The adversary (server) receives the local gradients after local training and updates the global model. However, it may attempt to perform gradient leakage attacks on the shared gradients to reconstruct clients' private training data. It initializes random dummy data and puts the dummy data into the global model. Then the corresponding dummy gradients are calculated. The adversary optimizes the distance-based loss function between the dummy gradients and the shared gradients from private data to iteratively update dummy data to approximate the actual private data. 
We found that the privacy attack converges fast with only a small number of iterations, far from reaching the pre-defined total number of iterations, leaving a huge potential to optimize the privacy attack efficiency.

\section{Methodology}

\subsection{Attack Method}

The goal of the privacy attack is to reconstruct private training data sample $x \in X$ and label $y \in Y$ from the shared gradients. The gradient $\nabla_{W} L$ is calculated and shared after training the local model $F_{W}$ on $(x, y)$.
The model parameters $W$ are optimized using the training loss function $L$. We first randomly initialize the dummy input data $x'$. Then we predict the label $y'$ to recover a private training sample $x$. We follow the method suggested in~\cite{zhao2020idlg} for finding the true label. The sign of the gradient corresponding to the true label logit in the output layer is opposite to the sign of the gradient of other output layer logits. For a neural network with $l$ layers, the label $y'$ can be predicted in Formula~\ref{formula:label-prediction-iDLG}.

\begin{equation}
y'=i \quad s.t. \quad \nabla W^i_l.\nabla W^j_l\leq 0\ , \forall j \neq i
\label{formula:label-prediction-iDLG}
\end{equation}
where $\nabla W^i_l$ is the gradient of the weight $W_l$ on the $i^{th}$ logit of the output layer, corresponding to label $i$.

Then we calculate the dummy gradients as Formula~\ref{formula:dummy-gradients-iDLG} shows.
\begin{equation}
\nabla W' = \nabla_{W} L(F_{W}(x'),y')  
\label{formula:dummy-gradients-iDLG}
\end{equation}

Upon obtaining the dummy gradients, the distance between the actual gradient and the dummy gradient is iteratively minimized to gradually update the dummy data to approach the real private data. 
A distance-based loss function is utilized to facilitate this process. In this study, we use the Euclidean distance between $\nabla W$ (shared gradients on private data) and $\nabla W'$ (dummy gradients) as the distance function $Dist$ in Formula~\ref{formula:distance-function-iDLG}.

\begin{equation}
Dist(\nabla W',\nabla W) = || \nabla W'-\nabla W||^2
\label{formula:distance-function-iDLG}
\end{equation}

The following objective (Formula~\ref{formula:attack-optimization-goal-iDLG}) is adopted to recover the private training data.

\begin{equation}
\underset{x',y'}{min \:} Dist(\nabla W',\nabla W)
\label{formula:attack-optimization-goal-iDLG}
\end{equation}

Here, the dummy data can be updated by using an optimizer, such as SGD with a learning rate $\eta$, to reconstruct the private training sample $x$ as shown in Formula~\ref{formula:attack-sgd-iDLG}.

\begin{equation}
x' = x' - \eta\frac{\partial Dist(\nabla W',\nabla W)}{\partial x'}
\label{formula:attack-sgd-iDLG}
\end{equation}

\subsection{Overview of EPAFL}

\begin{figure}[h!]
\centering
  \begin{subfigure}[b]{0.495\linewidth}
    \includegraphics[width=1.0\linewidth]{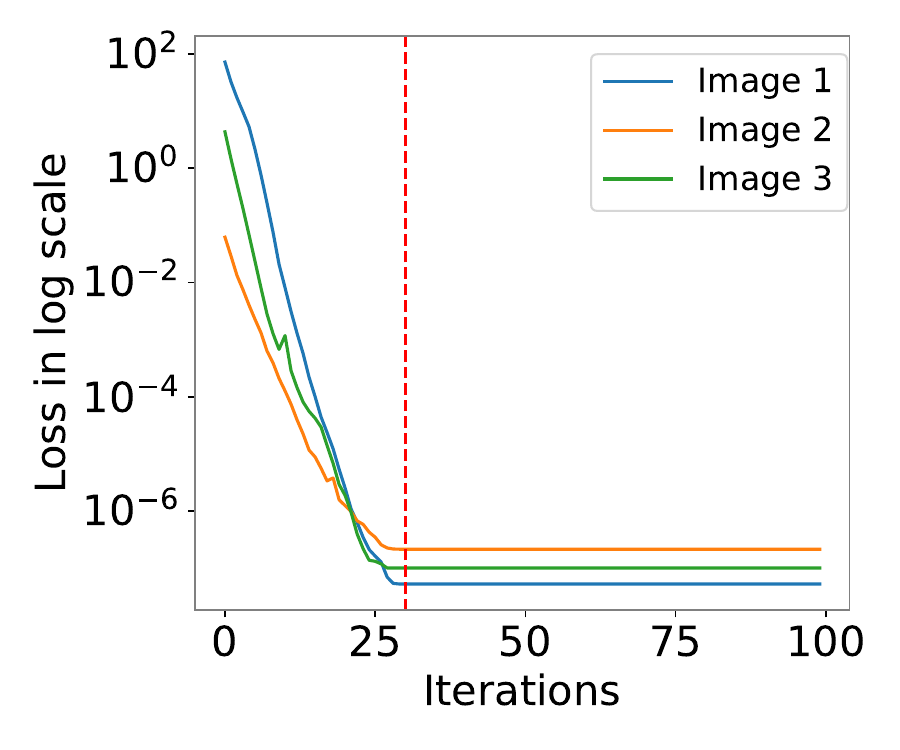}
  \end{subfigure}
  \begin{subfigure}[b]{0.495\linewidth}   \includegraphics[width=1.0\linewidth]{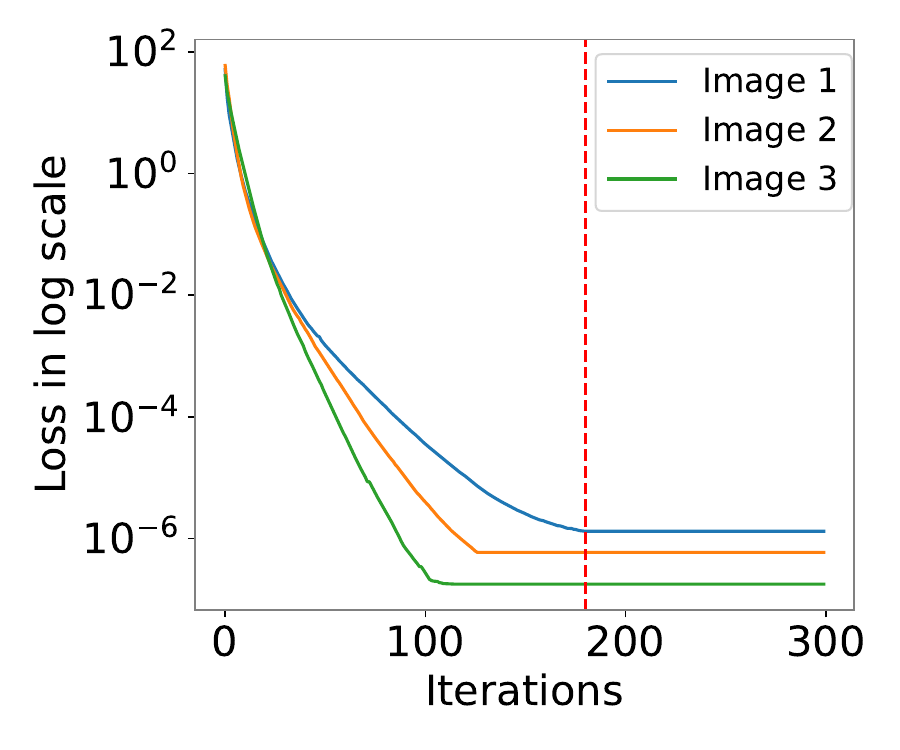}
  \end{subfigure}
  \caption{Loss vs. Iterations for Three MNIST (left) and Three CIFAR-10 (right) Samples}
  \label{fig:loss_curve}
\end{figure}

\begin{algorithm}[tb]
    \caption{Privacy Attack with Hybrid Early Stopping}
    \label{alg:algorithm-hybrid}
    \textbf{Input}: $F(x,W)$: Differentiable model, $\nabla W$: Gradients obtained from local training on private data $(x,y)$, $M$: Monitored metrics distance loss, $N$: Maximum number of iterations, 
    $T$: Threshold, $P$: Patience, $\eta$: Attack learning rate\\
    \textbf{Output}: Reconstructed training sample $(x',y')$\\
    \textbf{Initialize}: \mbox{$x' \leftarrow \mathcal{N}(0,1)$}
    \begin{algorithmic}[1] 
        \STATE Calculate the dummy label:
        \STATE $y' \leftarrow j \quad s.t. \quad \nabla W^j_l.\nabla W^k_l\leq 0\ , \forall k \neq j$ 
        \FOR{$i$ from $1$ to $N$}
        \STATE Calculate the dummy gradient:
        \STATE $\nabla W' \leftarrow \nabla_{W} L(F_{W}(x'),y')$
        \STATE Calculate the distance between dummy gradient $\nabla W'$ and actual gradient $\nabla W$:
        \STATE $Dist(\nabla W',\nabla W) \leftarrow || \nabla W'-\nabla W||^2$
        \STATE Update dummy data:
        \STATE $x' \leftarrow x' - \eta\frac{\partial Dist(\nabla W',\nabla W)}{\partial x'}$
        \STATE Apply threshold-based early stopping:
        \STATE $early\_stop \leftarrow Is\text{-}Below\text{-}Threshold(M, T)$
        \IF {$early\_stop == true$}
        \STATE $I \leftarrow i$
        \STATE \textbf{break}
        \ENDIF
        \STATE Apply plateau-based early stopping:
        \STATE $early\_stop \leftarrow Is\text{-}Trapped\text{-}On\text{-}Plateau(M, P)$
        \IF {$early\_stop == true$}
        \STATE $I \leftarrow i$
        \STATE \textbf{break}
        \ENDIF
        \ENDFOR
        \STATE \textbf{return} $(x',y')$
    \end{algorithmic}
\end{algorithm}

\begin{figure}
\centering  \includegraphics[width=0.9\linewidth]{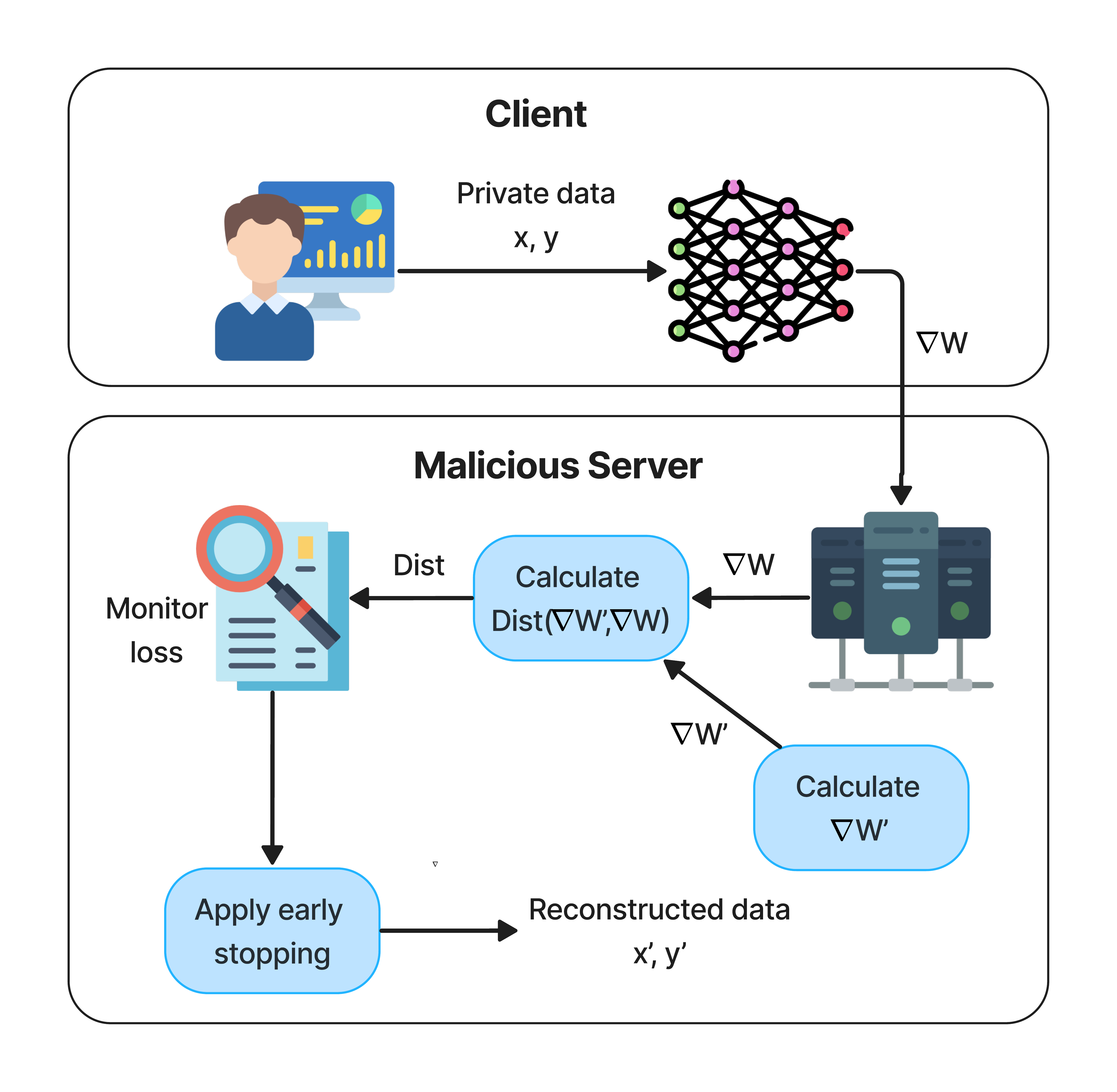}
  \caption{Architecture Overview of EPAFL}
  \label{fig:EPAFL}
  \vspace{-3mm}
\end{figure}

The main objective of our EPAFL framework is to enhance the efficiency of privacy attacks so that private data can be reconstructed in fewer iterations and in less execution time. 
Figure~\ref{fig:attack-visualize} visualizes the privacy attack results on four image samples. Here, we observe that private data can be successfully reconstructed in much fewer iterations than the pre-defined 100 iterations.
We also plot the loss curves in Figure~\ref{fig:loss_curve}. We find that the loss will stop improving after a certain number of iterations as marked by the red dashed lines. This indicates that the privacy attack process can be terminated before the pre-defined number of iterations. The key challenge here is how to determine when to stop the privacy attack, which motivates us to investigate early-stopping techniques in this study. Early-stopping techniques hold the potential to terminate the privacy attack procedure as soon as the loss stops improving and reduce the execution time costs of privacy attacks. 
Figure~\ref{fig:EPAFL} presents the architecture overview of the proposed framework EPAFL, which leverages status monitors, e.g., monitoring data reconstruction loss, and applies various early stopping techniques to significantly enhance the Efficiency of Privacy Attacks in FL.

We propose an efficient privacy attack method in Algorithm~\ref{alg:algorithm-hybrid}. Here, various early stopping strategies are employed based on the distance loss to establish effective conditions for terminating the privacy attack early without compromising the quality of reconstructed data. EPAFL monitors the loss values to determine whether to run the privacy attack for the pre-defined fixed number of iterations $N$ or stop the attack process after the required number of iterations $I$.  
In this section, we primarily describe three different early stopping techniques that we have investigated using EPAFL: (1) threshold-based technique, (2) plateau-based technique, and (3) their combination. 

\subsubsection{Threshold-based Early Stopping}

An intuitive method for early stopping is to terminate the privacy attack procedure when the reconstruction loss is below a threshold. 
The function $Is\text{-}Below\text{-}Threshold(M, T)$ takes two arguments: the monitored metric value $M$, such as the loss value, and a threshold value $T$, and checks if the monitored metric is below the predefined threshold $T$. If $M < T$, the privacy attack will be terminated. Otherwise, the privacy attack will continue until reaching the pre-defined number of iterations $N$. The threshold value $T$ should be chosen carefully as very small values of $T$ may not stop the privacy attack procedure timely. On the other hand, a very large threshold ($T$) may reduce the execution time at the cost of poor quality of the reconstructed data.

\subsubsection{Plateau-based Early Stopping}

\begin{algorithm}[tb]
    \caption{Is-Trapped-On-Plateau}
    \label{alg:algorithm-plateau-based}
    \textbf{procedure} Is-Trapped-On-Plateau($M, P$)\\
    \textbf{Input}: $M$: Monitored metrics, $P$: Patience\\
    \textbf{Output}: Early stop decision\\
    \textbf{Initialize:}\\
    $wait \leftarrow 0$\\
    $plateau\_start \leftarrow false$\\
    $early\_stop \leftarrow false$ \\
    $M_{best} \leftarrow \infty$
    
    \begin{algorithmic}[1] 
        \FOR{$i$ from $1$ to $P$}
        \IF {$M_{i} < M_{best}$}
        \STATE $M_{best} \leftarrow M_{i}$ 
        \STATE $wait \leftarrow 0$
        \STATE $plateau\_start \leftarrow false$
        \ELSE
        \STATE $wait \leftarrow wait + 1$
        \STATE $plateau\_start \leftarrow true$
        \ENDIF
        \ENDFOR
        \IF {$wait == P$ and $plateau\_start == true$}
        \STATE $early\_stop \leftarrow true$
        \ENDIF
        \STATE \textbf{return} $early\_stop$
        \STATE \textbf{end procedure}
    \end{algorithmic}
\end{algorithm}

Plateau detection is another effective technique for applying early stopping and enhancing the efficiency of deep learning training~\cite{reducelronplateau,wu2019demystifying,wu2023selecting}. 
Plateau-based early stopping monitors the loss value for multiple consecutive iterations. Given that the loss values tend to oscillate, the loss value may not be stable during the privacy attack procedure. Hence, we set a patience value $P$ which indicates how many consecutive iterations the loss will be monitored for before stopping the privacy attack process.
If there is no improvement in the loss for several consecutive iterations, the privacy attack will be terminated. 
Algorithm~\ref{alg:algorithm-plateau-based} shows the procedure for the plateau-based early-stopping. It takes two arguments:  the monitored metric value $M$ (e.g., the loss value) and patience $P$. When the loss $M_{i}$ in iteration $i$ becomes larger than or equal to the lowest loss $M_{best}$, the algorithm starts monitoring the loss for $P$ successive iterations, e.g., for 10 iterations, and checks if the loss is trapped on a plateau. If the loss stagnates, indicating no improvements in data reconstruction, the privacy attack will be terminated. Otherwise, the privacy attack will continue until the next detection of loss plateau or reaching the pre-defined number of iterations.
The patience value $P$ should be chosen carefully as very small values of $P$ may stop the privacy attack too early and impair the data reconstruction quality. On the other hand, very large values of $P$ may result in high execution costs.

\subsubsection{Hybrid Early Stopping}
In our experiments, we observe that for unsuccessful privacy attacks, the loss value often stagnates at a higher value than the threshold, which may not be terminated early by only using the threshold-based method. On the other hand, when the loss is already below the pre-defined threshold with a loss very close to 0, the plateau-based method still requires $P$ iterations before terminating the privacy attack process, leading to unnecessary execution time costs.
Hence, we combine both threshold-based and plateau-based early-stopping techniques to form a hybrid method. This hybrid early-stopping method keeps track of the loss and terminates the privacy attack procedure when any early-stopping condition is met. 
Algorithm~\ref{alg:algorithm-hybrid} shows this hybrid early stopping technique. Integrating two early-stopping techniques can harness the advantages of both threshold-based and plateau-based early-stopping techniques and significantly enhance privacy attack efficiency. 

\section{Experimental Analysis}

\subsection{Experimental Setup}

We implement EPAFL with PyTorch and perform experiments on two benchmark image datasets, MNIST dataset~\cite{mnistlenet} of image size 28$\times$28 and CIFAR-10 dataset~\cite{cifar10-100} of image size 32$\times$32. We use the following configurations for performing the experiments. The learning rate $\eta$=1.0 for the LBFGS optimizer. We utilize an untrained LeNet~\cite{mnistlenet} to conduct the experiments. 
To initialize the weights of the neural network, we follow a uniform distribution with values ranging from -0.5 to 0.5. The dummy data are generated using a Gaussian distribution. We randomly choose a subset of 100 samples from the dataset and perform gradient leakage attacks on those samples. 
The maximum number of iterations $N$ is set as 300, implying that if the reconstruction process does not stop earlier, the privacy attack will run for 300 iterations for each sample.

\begin{table*}[!h]
\centering
\begin{tabular}{|c|c|c|c|c|c|c|c|c|c|}
\hline
                          Dataset & Threshold & ASR  & \begin{tabular}[c]{@{}c@{}}MSE\\ (Avg.)\end{tabular}       & \begin{tabular}[c]{@{}c@{}}SSIM\\ (Avg.)\end{tabular}   & \begin{tabular}[c]{@{}c@{}}Reconstruction \\ Time (s)\end{tabular} & \begin{tabular}[c]{@{}c@{}}Iterations \\ (Max)\end{tabular} & \begin{tabular}[c]{@{}c@{}}Iterations \\ (Min)\end{tabular} & \begin{tabular}[c]{@{}c@{}}Iterations \\ (Avg.)\end{tabular} & \begin{tabular}[c]{@{}c@{}}Iterations \\ (SD)\end{tabular} \\ \hline
\multirow{4}{*}{MNIST}    & 0.001     & 0.65 & 0.001     & 0.9579 & \textbf{125.26}                                                             & 18                                                          & \textbf{5}                                                           & \textbf{7.4}                                                          & \textbf{2.154}                                                      \\ \cline{2-10} 
                          & 0.0001    & 0.68 & 0.0001    & 0.9694 & 183.84                                                             & \textbf{15}                                                          & 6                                                           & 9.44                                                         & 2.165                                                      \\ \cline{2-10} 
                          & 0.00001   & 0.74 & \textbf{2.955e-05} & 0.977  & 281.85                                                             & 24                                                          & 8                                                           & 12.72                                                        & 3.411                                                      \\ \cline{2-10} 
                          & 0.000001  & \textbf{0.79} & 3.576e-05 & \textbf{0.9787} & 371.03                                                             & 34                                                          & 9                                                           & 15.15                                                        & 4.984                                                      \\ \hline
\multirow{4}{*}{CIFAR-10} & 0.001     & 0.14 & 0.0054    & 0.9354 & \textbf{234.65}                                                             & \textbf{71}                                                          & \textbf{37}                                                         & \textbf{49}                                                           & \textbf{9.769}                                                      \\ \cline{2-10} 
                          & 0.0001    & 0.55 & 0.001     & 0.9674 & 1525.31                                                            & 124                                                         & 54                                                          & 78.22                                                        & 16.466                                                     \\ \cline{2-10} 
                          & 0.00001   & \textbf{0.74} & 0.0002    & 0.9874 & 3050.43                                                            & 284                                                         & 66                                                          & 114.88                                                       & 31.76                                                      \\ \cline{2-10} 
                          & 0.000001  & 0.72 & \textbf{4.434e-05} & \textbf{0.9956} & 3763.76                                                            & 300                                                         & 82                                                          & 180.42                                                       & 81.833                                                      \\ \hline
\end{tabular}
\caption{Privacy Attack Performance under Different Thresholds (Threshold-based Early Stopping)}
\label{tab:threshold}
\end{table*}

\begin{table*}[]
\centering
\begin{tabular}{|c|c|c|c|c|c|c|c|c|c|}
\hline
Dataset                   & Patience & ASR  & \begin{tabular}[c]{@{}c@{}}MSE\\ (Avg.)\end{tabular}       & \begin{tabular}[c]{@{}c@{}}SSIM\\ (Avg.)\end{tabular}   & \begin{tabular}[c]{@{}c@{}}Reconstruction \\ Time (s)\end{tabular} & \begin{tabular}[c]{@{}c@{}}Iterations\\ (Max)\end{tabular} & \begin{tabular}[c]{@{}c@{}}Iterations\\ (Min)\end{tabular} & \begin{tabular}[c]{@{}c@{}}Iterations\\ (Avg.)\end{tabular} & \begin{tabular}[c]{@{}c@{}}Iterations\\ (SD)\end{tabular} \\ \hline
\multirow{3}{*}{MNIST}    & 5        & 0.82 & \textbf{1.833e-07} & 0.9823 & \textbf{313.68}                                                             & \textbf{45}                                                        & \textbf{20}                                                         & \textbf{26.44}                                                      & \textbf{4.758}                                                     \\ \cline{2-10} 
                          & 10       & 0.80 & 9.371e-07 & \textbf{0.983}  & 327.91                                                             & 95                                                         & 23                                                         & 32.41                                                       & 8.502                                                     \\ \cline{2-10} 
                          & 15       & \textbf{0.84} & 3.526e-05 & 0.9824 & 361.38                                                             & 63                                                         & 28                                                         & 37.35                                                       & 6.062                                                     \\ \hline
\multirow{3}{*}{CIFAR-10} & 5        & 0.69 & \textbf{2.745e-05} & \textbf{0.997}  & \textbf{2598.12}                                                            & \textbf{280}                                                        & 106                                                        & \textbf{175.04}                                                      & \textbf{41.19}                                                     \\ \cline{2-10} 
                          & 10       & 0.69 & 4.78e-05  & 0.9956 & 2974.91                                                            & 300                                                        & \textbf{105}                                                       & 180.01                                                      & 49.35                                                     \\ \cline{2-10} 
                          & 15       & \textbf{0.71} & 4.139e-05 & 0.996  & 4031.03                                                            & 300                                                        & 110                                                        & 185.51                                                      & 49.87                                                     \\ \hline
\end{tabular}
\caption{Privacy Attack Performance under Different Patience Values (Plateau-based Early Stopping)}
\label{tab:plateau}
\end{table*}

\subsection{Threshold-based Early Stopping}

First, we perform experiments by setting different values as the threshold for the loss. In this case, the privacy attack will be stopped after the loss goes below a certain threshold $T$. In the first step, we take a small number of samples and perform gradient attacks on them to empirically choose the appropriate threshold values for the experiments. 
To study the impact of the threshold-based early stopping on Attack Success Rate (ASR), we check whether the SSIM (Structural Similarity Index Measure) is above 0.9. In this way, we ensure that the reconstruction quality is consistently maintained. A higher threshold results in a lower SSIM as the privacy attacks stop earlier before fully reconstructing the private training samples. On the other hand, a lower threshold results in higher computational costs. The privacy attack continues until the loss is very small, where the reconstruction time and number of iterations increase. Based on this observation, we choose four different threshold values to perform the experiments. 

Table~\ref{tab:threshold} shows the impact of varying the threshold values on the privacy attacks. We make three interesting observations. 
{\it First,} the attack success rate (ASR) is affected by the specific threshold value. We find that the ASR increases as the threshold decreases. For the MNIST dataset, the lowest ASR is 65\%. However, the lowest ASR for CIFAR-10 is 14\% when the threshold is set to 0.001. We attribute this observation to the high complexity and rich content of the CIFAR-10 dataset. As a large number of iterations are required to reconstruct CIFAR-10 samples, stopping too early may result in lower ASR. 
{\it Second,} the MSE (Mean Squared Error, computed as an average over all samples) is quite low for all thresholds, which indicates that the distance between actual and dummy data can be minimized even after applying threshold-based early stopping. The average SSIM is above 0.93 for all thresholds, implying that good reconstruction quality can be achieved within a small number of iterations. 
{\it Third,} applying a threshold for early stopping heavily influences the reconstruction time and number of iterations. For example, 79\% of samples from the MNIST dataset can be successfully reconstructed in only 371.03 seconds and 15 iterations on average. The overall reconstruction time is very low, ranging from 125.26 to 371.03 seconds for MNIST and 234.65 to 3763.76 seconds for CIFAR-10. The results suggest that selecting a proper threshold can significantly improve privacy attack efficiency by reducing computational costs.

\begin{table*}[!h]
\centering
\begin{tabular}{|c|c|c|c|c|c|c|c|c|c|}
\hline
Patience            & Threshold & ASR  & \begin{tabular}[c]{@{}c@{}}MSE\\ (Avg.)\end{tabular} & \begin{tabular}[c]{@{}c@{}}SSIM\\ (Avg.)\end{tabular} & \begin{tabular}[c]{@{}c@{}}Reconstruction \\ Time (s)\end{tabular} & \begin{tabular}[c]{@{}c@{}}Iterations\\ (Max)\end{tabular} & \begin{tabular}[c]{@{}c@{}}Iterations\\ (Min)\end{tabular} & \begin{tabular}[c]{@{}c@{}}Iterations\\ (Avg.)\end{tabular} & \begin{tabular}[c]{@{}c@{}}Iterations\\ (SD)\end{tabular} \\ \hline
\multirow{4}{*}{5}  & 0.001     & 0.60 & 0.001                                                & 0.9596                                                & 126.64                                                             & 18                                                         & 5                                                          & 7.62                                                        & 2.346                                                     \\ \cline{2-10} 
                    & 0.0001    & 0.67 & 0.0002                                               & 0.9686                                                & 185.75                                                             & 36                                                         & 6                                                          & 9.70                                                        & 3.879                                                     \\ \cline{2-10} 
                    & 0.00001   & 0.77 & 1.755e-05                                            & 0.9738                                                & 287.27                                                             & 33                                                         & 7                                                          & 12.43                                                       & 3.859                                                     \\ \cline{2-10} 
                    & 0.000001  & 0.80 & 3.692e-05                                            & 0.9776                                                & 369.20                                                             & 35                                                         & 9                                                          & 15.29                                                       & 5.146                                                     \\ \hline
\multirow{4}{*}{10} & 0.001     & 0.62 & 0.0009                                               & 0.96                                                  & 115.75                                                             & 13                                                         & 5                                                          & 7.21                                                        & 1.815                                                     \\ \cline{2-10} 
                    & 0.0001    & 0.70 & 0.0001                                               & 0.9668                                                & 191.88                                                             & 37                                                         & 6                                                          & 9.94                                                        & 4.157                                                     \\ \cline{2-10} 
                    & 0.00001   & 0.76 & 3.438e-05                                            & 0.9745                                                & 285.23                                                             & 26                                                         & 8                                                          & 12.62                                                       & 3.479                                                     \\ \cline{2-10} 
                    & 0.000001  & 0.76 & 6.235e-05                                            & 0.9799                                                & 402.05                                                             & 59                                                         & 9                                                          & 17.17                                                       & 7.116                                                     \\ \hline
\multirow{4}{*}{15} & 0.001     & 0.62 & 0.0008                                               & 0.9596                                                & 114.35                                                             & 16                                                         & 5                                                          & 7.24                                                        & 1.775                                                     \\ \cline{2-10} 
                    & 0.0001    & 0.66 & 0.001                                                & 0.966                                                 & 228.93                                                             & 77                                                         & 6                                                          & 12.24                                                       & 10.07                                                     \\ \cline{2-10} 
                    & 0.00001   & \textbf{0.80} & 3.736e-05                                            & 0.9715                                                 & \textbf{274.73}                                                             & 20                                                         & 8                                                          & 12.04                                                       & 2.799                                                     \\ \cline{2-10} 
                    & 0.000001  & 0.83 & 2.309e-05                                            & 0.9786                                                & 389.58                                                             & 38                                                         & 9                                                          & 15.58                                                       & 5.109                                                     \\ \hline
\end{tabular}
\caption{Privacy Attack Performance under Different Thresholds and Patience Values (Hybrid Early Stopping, MNIST)}
\label{tab:plateau-threshold-mnist}
\end{table*}

\begin{table*}[!h]
\centering
\begin{tabular}{|c|c|c|c|c|c|c|c|c|c|}
\hline
Patience            & \multicolumn{1}{l|}{Threshold} & ASR  & \begin{tabular}[c]{@{}c@{}}MSE\\ (Avg.)\end{tabular} & \begin{tabular}[c]{@{}c@{}}SSIM\\ (Avg.)\end{tabular} & \begin{tabular}[c]{@{}c@{}}Reconstruction \\ Time (s)\end{tabular} & \begin{tabular}[c]{@{}c@{}}Iterations\\ (Max)\end{tabular} & \begin{tabular}[c]{@{}c@{}}Iterations\\ (Min)\end{tabular} & \begin{tabular}[c]{@{}c@{}}Iterations\\ (Avg.)\end{tabular} & \begin{tabular}[c]{@{}c@{}}Iterations\\ (SD)\end{tabular} \\ \hline
\multirow{4}{*}{5}  & 0.001                          & 0.16 & 0.006                                                & 0.934                                                 & 289.34                                                             & 88                                                         & 38                                                         & 50.31                                                       & 12.128                                                    \\ \cline{2-10} 
                    & 0.0001                         & 0.57 & 0.001                                                & 0.9626                                                & 1624.35                                                            & 300                                                        & 51                                                         & 84.96                                                       & 36.559                                                    \\ \cline{2-10} 
                    & 0.00001                        & \textbf{0.70} & 0.0002                                               & 0.9898                                                & \textbf{2895.69}                                                            & 230                                                        & 71                                                         & 120.66                                                      & 40.117                                                    \\ \cline{2-10} 
                    & 0.000001                       & 0.71 & 4.706e-05                                            & 0.9955                                                & 3766.06                                                            & 300                                                        & 81                                                         & 163.46                                                      & 60.82                                                     \\ \hline
\multirow{4}{*}{10} & 0.001                          & 0.13 & 0.005                                                & 0.9386                                                & 193.34                                                             & 62                                                         & 37                                                         & 47.54                                                       & 8.044                                                     \\ \cline{2-10} 
                    & 0.0001                         & 0.55 & 0.001                                                & 0.9642                                                & 1422.47                                                            & 123                                                        & 53                                                         & 79.22                                                       & 15.161                                                    \\ \cline{2-10} 
                    & 0.00001                        & 0.75 & 0.0002                                               & 0.9873                                                & 2926.45                                                            & 248                                                        & 60                                                         & 117.71                                                      & 36.764                                                    \\ \cline{2-10} 
                    & 0.000001                       & 0.72 & 6.909e-05                                            & 0.9968                                                & 3754.77                                                            & 300                                                        & 90                                                         & 162.28                                                      & 51.827                                                    \\ \hline
\multirow{4}{*}{15} & 0.001                          & 0.16 & 0.005                                                & 0.9318                                                & 241.86                                                             & 66                                                         & 37                                                         & 47.06                                                       & 7.521                                                     \\ \cline{2-10} 
                    & 0.0001                         & 0.58 & 0.001                                                & 0.9672                                                & 1581.15                                                            & 166                                                        & 55                                                         & 80.66                                                       & 20.529                                                    \\ \cline{2-10} 
                    & 0.00001                        & 0.66 & 0.0002                                               & 0.9906                                                & 2718.70                                                            & 300                                                        & 67                                                         & 121.56                                                      & 48.399                                                    \\ \cline{2-10} 
                    & 0.000001                       & 0.80 & 5.009e-05                                            & 0.9943                                                & 4778.28                                                            & 300                                                        & 76                                                         & 166.35                                                      & 54.68                                                     \\ \hline
\end{tabular}
\caption{Privacy Attack Performance under Different Thresholds and Patience Values (Hybrid Early Stopping, CIFAR-10)}
\label{tab:plateau-threshold-cifar10}
\end{table*}

\subsection{Plateau-based Early Stopping}

In this set of experiments, we evaluate the performance of the plateau-based early stopping by detecting the loss plateau. 
If the loss does not improve for a few consecutive iterations (specified by patience $P$), we terminate the reconstruction process early to reduce the computational costs. 
We choose \{5, 10, 15\} as patience values to investigate the impact of this early-stopping technique. 

Table~\ref{tab:plateau} presents the impact of different patience values on the computational costs and performance of the privacy attacks. We highlight two interesting observations below. 
{\it First,} the plateau-based early stopping performs better than the threshold-based technique (in Table~\ref{tab:threshold}) in terms of ASR and SSIM. For the MNIST dataset, the ASR is above 80\% for all patience values, which indicates that this technique works well for reconstructing most of the samples. We observe that the ASR is significantly higher for both datasets when patience is set to 15. The SSIM is also very high for different patience values. For example, the best average SSIM of the successfully reconstructed CIFAR-10 samples is 0.997 which implies that high-quality reconstructed data can be obtained even after employing this plateau-based early stopping method. 
{\it Second,} a lower patience $P$ can reduce the computational costs. In Table~\ref{tab:plateau}, the lowest data reconstruction cost is achieved by $P$=5 in terms of the reconstruction time and number of iterations. For MNIST, it takes only 313.68 seconds in total to reveal 82\% of samples with an average of 26 iterations per sample. For CIFAR-10, it requires 175 iterations on average to successfully reconstruct 69\% of the samples with a total execution time of 2598.12 seconds. The results show that the computational costs can be significantly reduced by stopping the privacy attack when the loss reaches a plateau.

\begin{table*}[!th]
\centering
\begin{tabular}{|c|ccc|ccc|}
\hline
\multirow{2}{*}{Metrics}       & \multicolumn{3}{c|}{MNIST}                                             & \multicolumn{3}{c|}{CIFAR-10}                                               \\ \cline{2-7} 
                        & \multicolumn{1}{c|}{DLG}    & \multicolumn{1}{c|}{iDLG}    & EPAFL      & \multicolumn{1}{c|}{DLG}       & \multicolumn{1}{c|}{iDLG}      & EPAFL      \\ \hline
ASR                     & \multicolumn{1}{c|}{0.66}   & \multicolumn{1}{c|}{0.79}    & \textbf{0.80}      & \multicolumn{1}{c|}{0.69}      & \multicolumn{1}{c|}{0.72}      & \textbf{0.75}      \\ \hline
MSE (Avg.)              & \multicolumn{1}{c|}{0.0004} & \multicolumn{1}{c|}{0.0003}  & {\textbf{3.736e-05}} & \multicolumn{1}{c|}{\textbf{4.194e-05}} & \multicolumn{1}{c|}{4.211e-05} & 0.0002 \\ \hline
SSIM (Avg.)             & \multicolumn{1}{c|}{\textbf{0.9839}} & \multicolumn{1}{c|}{0.9824}  & 0.9715    & \multicolumn{1}{c|}{\textbf{0.9961}}    & \multicolumn{1}{c|}{0.996}     & 0.9873    \\ \hline
Reconstruction Time (s) & \multicolumn{1}{c|}{1029.59} & \multicolumn{1}{c|}{1022.07} & \textbf{274.73}    & \multicolumn{1}{c|}{4797.01}   & \multicolumn{1}{c|}{4207.33}   & \textbf{2926.45}   \\ \hline
\end{tabular}
\caption{Comparison of Our Hybrid Early Stopping with Representative Privacy Attacks}
\label{tab:baseline-comparison}
\vspace{-3mm}
\end{table*}

\subsection{Hybrid Early Stopping}

We then evaluate the hybrid early stopping mechanism, which integrates both threshold-based and plateau-based early stopping techniques. We vary both the patience and threshold values to study their combined effects on privacy attack performance.
We present the experimental results in Table~\ref{tab:plateau-threshold-mnist} for MNIST and Table~\ref{tab:plateau-threshold-cifar10} for CIFAR-10 respectively. We highlight two interesting observations. 
{\it First,} the hybrid early stopping method can effectively outperform the individual threshold-based or plateau-based early stopping techniques by achieving comparable ASR at a reduced execution cost. For MNIST, the hybrid early stopping can achieve over 80\% ASR with only 274.73 seconds, which is much faster than the plateau-based (313.68 seconds) early stopping technique. Similarly, for CIFAR-10, the hybrid early stopping technique can achieve over 70\% ASR with only 2895.69 seconds, significantly outperforming the threshold-based (3763.76 seconds) and plateau-based (4031.03 seconds) early stopping techniques.
{\it Second,} both the patience and threshold values have high impact on the data reconstruction quality and costs. For CIFAR-10, the lowest reconstruction time is 193.34 seconds with threshold $T$=0.001 and patience $P$=10. However, the ASR is only 13\% in this case. A higher ASR (75\%) can be achieved by decreasing the threshold to 0.00001 and also outperforming the individual threshold-based or plateau-based methods in terms of reconstruction time. An interesting observation is that if we set $T$=0.000001 and $P$=15 for CIFAR-10, the ASR (80\%) becomes quite higher but at the cost of a higher reconstruction time. 
This highlights the necessity of choosing a suitable combination of threshold and patience to maintain a higher attack success rate at lower computational costs.

\subsection{Comparison to Other Privacy Attacks}

We next compare our hybrid early stopping with two representative privacy attacks: DLG \cite{zhu2019deep} and iDLG \cite{zhao2020idlg}. 
Table~\ref{tab:baseline-comparison} presents the experimental results for comparing the three privacy attack methods. 
We primarily compare the data reconstruction quality metrics (ASR, MSE and SSIM) and the execution time. 
Three interesting observations should be highlighted.
{\it First,} we observe that our hybrid early stopping approach can outperform other privacy attack methods in terms of ASR. Here, we configure $P$=15 and $T$=0.00001 for MNIST (see Table~\ref{tab:plateau-threshold-mnist}) and $P$=10 and $T$=0.00001 (see Table~\ref{tab:plateau-threshold-cifar10}) for CIFAR-10. We attribute the enhanced ASR by EPAFL to the potential overfitting by DLG and iDLG for running the entire $N$ iterations.
{\it Second,} our early stopping method can achieve comparable MSE and SSIM to other privacy attack methods. In particular, our privacy attack method achieves the lowest average MSE for MNIST.  
{\it Third,} our early stopping method can significantly reduce the reconstruction time costs by over 73\% for MNIST and over 30\% for CIFAR-10.

\begin{figure}
\centering  \includegraphics[width=\linewidth]{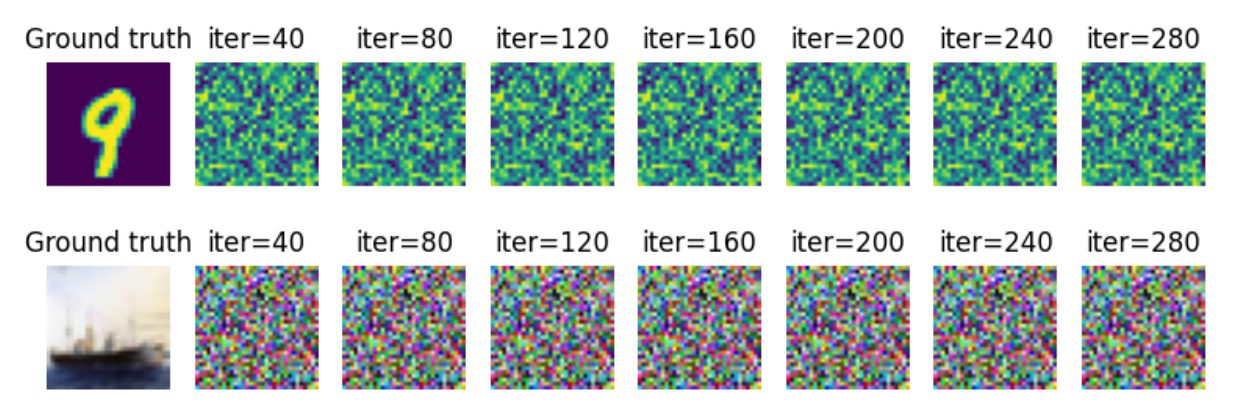}
  \caption{Failure Cases in Image Reconstruction}
  \label{fig:att-fail}
  \vspace{-3.5mm}
\end{figure}

\begin{figure}
\centering  \includegraphics[width=\linewidth]{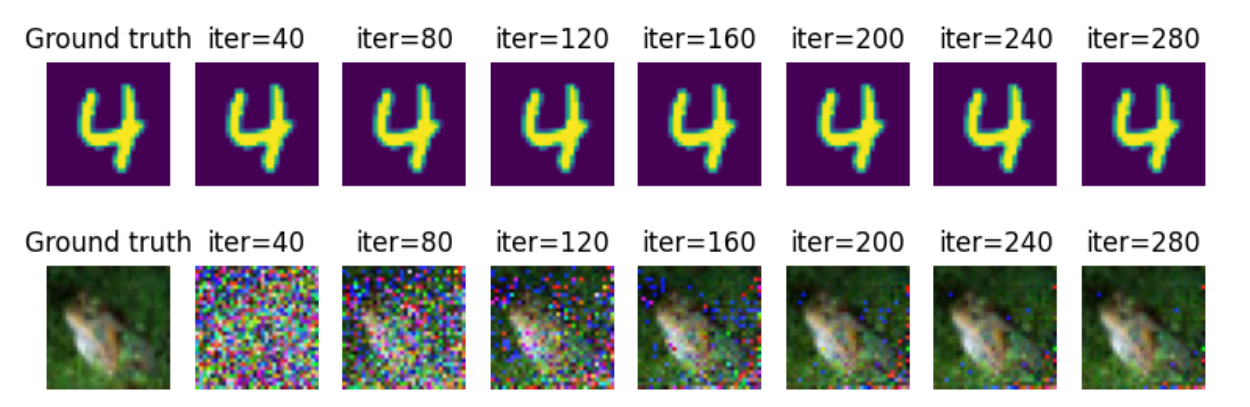}
  \caption{Successfully Reconstructed Images}
  \label{fig:att-succ}
  \vspace{-4.5mm}
\end{figure}

\subsection{Visualization}

We present two different cases in privacy attacks: (1) attack failures in Figure~\ref{fig:att-fail} and (2) successful attacks in Figure~\ref{fig:att-succ}. For failure cases, the dummy data consistently exhibits noise across all iterations of the privacy attack. For the successful attacks, the private data can often be effectively recovered, significantly prior to the pre-defined 300 iterations, such as within 40 iterations for the first example in Figure~\ref{fig:att-succ}. These observations further demonstrate that stopping the privacy attack early can significantly enhance the efficiency of privacy attacks without impairing the attack effectiveness.

\section{Conclusion}

In this paper, we have identified a novel research problem of how to optimize the efficiency of privacy attacks in Federated Learning. We made three novel contributions.
{\it First,} we analyzed the performance of privacy attacks represented by gradient leakage attacks and identified the huge potential to optimize their efficiency.
{\it Second,} we proposed a holistic framework (EPAFL) to enhance the efficiency of privacy attacks by leveraging three early-stopping techniques.
{\it Third,} we conducted experiments on two benchmark vision datasets, which shows that our approach can effectively recover private training data at a much lower execution costs and maintain comparable attack success rates.
The proposed framework will provide fast evaluation of various privacy attacks and defense methods and facilitate efficient development of various defense strategies.
Our ongoing studies include (1) evaluating the EPAFL framework on other benchmark datasets, (2) supporting other privacy attack algorithms in EPAFL, and (3) investigating other techniques for improving the efficiency of both privacy attacks and defense methods.

{
    \small
    \bibliographystyle{ieeenat_fullname}
    \bibliography{reference}
}

\end{document}